Growth and physical properties of Ce(O,F)Sb(S,Se)$_2$ single crystals with site-selected chalcogen atoms

Masanori Nagao[a,*], Masashi Tanaka[b], Akira Miura[c], Miho Kitamura[d], Koji Horiba[d], Satoshi Watauchi[a], Yoshihiko Takano[e], Hiroshi Kumigashira[d], and Isao Tanaka[a]

[a]*University of Yamanashi, 7-32 Miyamae, Kofu, Yamanashi 400-8511, Japan*

[b]*Kyushu Institute of Technology, 1-1 Sensui-cho, Tobata, Kitakyushu, Fukuoka 804-8550, Japan*

[c]*Hokkaido University, Kita-13 Nishi-8, Kita-ku, Sapporo, Hokkaido 060-8628, Japan*

[d]*Photon Factory, Institute of Materials Structure Science, High Energy Accelerator Research Organization (KEK), 1-1 Oho, Tsukuba, Ibaraki 305-0801, Japan*

[e]*MANA, National Institute for Materials Science, 1-2-1 Sengen, Tsukuba, Ibaraki 305-0047, Japan*

*Corresponding Author

Masanori Nagao

Postal address: University of Yamanashi, Center for Crystal Science and Technology

Miyamae 7-32, Kofu, Yamanashi 400-8511, Japan

Telephone number: (+81)55-220-8610

Fax number: (+81)55-254-3035

E-mail address: mnagao@yamanashi.ac.jp

**Abstract**

Ce(O,F)Sb(S,Se)$_2$ single crystals were successfully grown using a CsCl/KCl flux method. The obtained crystals have a plate-like shape with the typical size of 1-2 mm and well-developed *ab*-plane, which enables X-ray single crystal structural analysis. The Ce(O,F)Sb(S,Se)$_2$ crystallizes in a monoclinic space group, *P*2$_1$/*m*, with lattice parameters of *a* = 4.121(7) Å, *b* = 4.109(7) Å, *c* = 13.233(15) Å, *β* = 97.94(7) °. It is composed of alternate stacking of Ce-(O,F) and Sb-SSe layers, and the Sb-SSe layer includes selective occupation of Se atoms in its in-plane site. The valence state of Ce is estimated to be Ce$^{3+}$ by X-ray absorption fine spectroscopy analysis. The single crystals show an insulating behavior, and a magnetic ordering around 6 K.

**Main text**

**1. Introduction**

$PnCh_2$-based compounds, $RnOPnCh_2$ ($Rn$ = rare earth elements, $Pn$ = Sb, Bi and $Ch$ = S, Se), provide a variety of research area, such as superconductivity [1-3], optoelectronic [4], thermoelectric [5] properties and superconductivity with magnetic ordering [6,7]. Moreover, they also become general insulator [8]. They crystallize with a layered structure composed of $PnCh_2$ layers and charge blocking $RnO$ layers, similar to which is found in Fe-based superconductors, $Rn$(O,F)FeAs ($Rn$: rare earth elements) [9,10]. There are two kinds of chalcogen sites; $Ch$1:in-plane, $Ch$2:out-plane, as is schematically illustrated in Figure 1.

Recently, a theoretical calculation suggests a possibility of topological insulator in an $SbSe_2$-based material, $LaOSbSe_2$ [11]. However, our trials to grow $RnOSbSe_2$ single crystals have not yet been succeeded.

In another way to construct SbSe planes, a site selectivity in $Ch$-sites might shed some light on our trials. The $Ch$-sites can be occupied either S or Se, resulting in a variety of $PnSSe$-based compounds [12], and a site selectively in the chalcogen atoms has been observed in La(O,F)BiSSe [13]; Se prefers to occupy the in-plane site while S locates out-plane site, in the continuous substitution between Se and S atoms in the $Ch$ sites. If the same thing would be happened in $Rn$(O,F)SbSSe, SbSe planes for topological insulators might be achieved.

In this paper, we have successfully grown Ce(O,F)Sb(S,Se)$_2$ single crystals using a CsCl/KCl flux method. The single crystals of Ce(O,F)Sb(S,Se)$_2$ were characterized by means of X-ray single crystal structural analysis, X-ray absorption fine spectroscopy, electrical/magnetic measurements and photoemission spectroscopy.

## 2. Experimental

Single crystals of Ce(O,F)Sb(S,Se)$_2$ were grown by a high-temperature flux method [14-16]. The raw materials of Ce, Sb$_2$O$_3$, SbF$_3$, Sb, Sb$_2$S$_3$, Sb$_2$Se$_3$, S, Se were weighed to have a nominal composition of Ce(O$_{1-u}$F$_u$)Sb(S$_{2-x}$Se$_x$) ($u$ = 0-0.5, $x$ = 1.0-1.5). The mixture of the raw materials (0.8 g) and CsCl/KCl flux (5.0 g) were ground by using a mortar, and then sealed into an evacuated quartz tube. The molar ratio of the CsCl/KCl flux was CsCl:KCl = 5:3. The quartz tube was heated at $T_{max}$:800-900 °C for 10 h, followed by cooling to 600 °C at a rate of 1 °C/h, then the sample was cooled down to room temperature in the furnace. The resulting quartz tube was opened in air, and the obtained materials were washed and filtered by distilled water in order to remove the CsCl/KCl flux. Also La(O,F)Sb(S,Se)$_2$ single crystals could be grown in the same method by changing the Ce elements to La in the raw materials.

The compositional ratio of the single crystals was evaluated by electron probe microanalysis (EPMA) associated with the observation of the microstructure by using scanning electron microscope (SEM) (JEOL, JXA-8200). The obtained compositional values were normalized using S+Se = 2.00, with Ce (or La) and Sb measured in a precision of two decimal places. After that, the F composition is normalized by the total F and O content. The characteristic X-ray signals of F-K$_\alpha$ (677 eV) and Ce-M$_\zeta$ (676 eV) are overlapped in EPMA analysis [17]. Then the incorporated F-values in Ce(O,F)Sb(S,Se)$_2$ were estimated by subtracting the value in the F-free CeOSb(S,Se)$_2$ ($u = 0$, $x = 1.0$) from the detected values.

Powder X-ray diffraction (XRD) patterns were measured by using Rigaku MultiFlex with CuK$\alpha$ radiation. Single crystal XRD structural analysis was carried out using a Rigaku Mercury CCD diffractometer with graphite monochromated MoK$\alpha$ radiation ($\lambda = 0.71072$ Å) (Rigaku, XtaLAB mini). The crystal structure was solved and refined by using the program SHELXT and SHELXL [18,19], respectively, in the WinGX software package [20].

Cerium valence state of the component was estimated by X-ray absorption fine spectroscopy (XAFS) analysis using an Aichi XAS beamline with a synchrotron X-ray radiation (BL5S1: Experimental No.201704041). For XAFS sample, the obtained single crystals were ground and mixed with boron nitride (BN) powder, followed by pressing into a pellet form with 4 mm diameter with total mass of around 28 mg.

Resistivity-temperature ($\rho$-$T$) characteristics of the obtained single crystal were measured by the standard four-probe method with a constant current density ($J$) mode using physical property measurement system (Quantum Design; PPMS DynaCool). The electrical terminals were made by silver paste. The temperature ($T$) dependence of magnetization ($M$) was measured by a superconducting quantum interference device (SQUID) magnetometer (MPMS, Quantum Design) under zero-field cooling (ZFC) and field cooling (FC) with an applied field ($H$) of 10 Oe parallel to the *c*-axis.

Photoemission spectroscopy (PES) measurements were performed at the BL-2A MUSASHI in the Photon Factory, KEK. The samples were cleaved *in situ* in ultra-high vacuum for obtaining fresh surfaces. All PES measurements were carried out at room temperature in order to avoid the charging effect. The binding energies were calibrated by $E_F$ of a gold plate.

### 3. Results and discussion

Figure 2 shows a typical SEM image for Ce(O,F)Sb(S,Se)$_2$ single crystals. The obtained single crystals had plate-like shape with 1.0-2.0 mm in size and 10-30 μm in thickness. Ce, O, F, Sb, S, Se elements were homogeneously detected in the obtained single crystals by qualitative analysis of EPMA. On the other hand, the flux components such as Cs, K, and Cl were not detected in the single crystals with a minimum sensitivity limit of 0.1 wt%. The atomic ratio of Ce:Sb in the single crystals were 1.06±0.03:0.97±0.03 which has almost same compared to the stoichiometry. Table I shows analytical compositions of the single crystals obtained with various heat treatment temperatures ($T_{max}$), starting material composition on the F contents ($u$) and Se contents ($x$). The Ce(O,F)Sb(S,Se)$_2$ single crystals were obtained from the nominal compositions with $u$ = 0-0.5 and $x$ = 1.0-1.5, except for $u$ = 0, $x$ = 1.5. The contents of F and Se in the obtained single crystals are smaller than the nominal composition. The Se substitution amount of S-site in Ce(O,F)Sb(S,Se)$_2$ single crystals increased with increasing the Se nominal composition. Single crystals with almost S:Se = 1:1 atomic ratio were grown from the nominal composition with $Rn$ = Ce, $u$ = 0.5, $x$ = 1.5.

The results in La(O,F)Sb(S,Se)$_2$ single crystals ($Rn$ = La) are also shown in Table I. In a nominal composition of $u$ = 0.5, $x$ = 1.0, F was not detected and the analytical Se content is only a half of the nominal composition. The crystals were not obtained in $u$ = 0.5, $x$ = 1.5.

Figure 3 shows the XRD pattern of a well-developed plane in the obtained Ce(O,F)Sb(S,Se)$_2$ single crystals from a starting powder with $u = 0.5$, $x = 1.5$. The appearance of only 00$l$ diffraction peaks indicates that the $ab$-plane is well-developed, which is similar to another SbS$_2$-based compound [8]. The X-ray single crystal structural analysis was carried out using the Ce(O,F)Sb(S,Se)$_2$ with $u = 0.5$, $x = 1.5$ single crystals. Details of the analysis and crystallographic parameters are listed in Table II and Table III. Although the convergence of $R_1$ values of the refinement is not enough as high as 14.41 % for $I > 4\sigma(I)$, the structural analysis tells us an important thing. The in-plane $Ch$1 site in Ce(O,F)Sb$Ch_2$ (Figure 1) was selectively occupied by Se atoms, otherwise the refinement had been diverged. The occupancy refinement was not deviated from 1.00 both in the S and Se atoms, and the result of Se is in good agreement with that found in the EPMA analysis. These findings suggest that the in-plane $Ch$1 sites of the Sb-SSe layer is perfectly occupied by Se atoms, supported by a similar site selectivity found in La(O,F)BiSSe single crystals [13]. In other words, the cleaved surface of Ce(O,F)Sb(S,Se)$_2$ with $u = 0.5$, $x = 1.5$ single crystals contains SbSe-plane, which is a candidate of above mentioned topological insulating layer. The bond valence sum of Ce atoms was calculated to be ~2.83, suggesting the absence of valence fluctuation fixed in Ce$^{3+}$.

The Ce valence state of Ce(O,F)Sb(S,Se)$_2$ single crystals was estimated by XAFS analysis. Figure 4 shows Ce L$_3$-edge absorption spectra of Ce(O,F)Sb(S,Se)$_2$ single crystals. Ce L$_3$-edge of Ce(O,F)Sb(S,Se)$_2$ single crystals showed the peak around 5724 eV, which can be assigned as Ce$^{3+}$, consistent with the another XAFS result for the trivalent electronic configuration (Ce$^{3+}$) [21]. There are no peaks at 5729 eV and 5736 eV assigned as tetravalent electronic configuration (Ce$^{4+}$). Therefore, valence state of Ce in the Ce(O,F)Sb(S,Se)$_2$ single crystals is only trivalent, suggesting that there are no valence fluctuation in good agreement with X-ray structural analysis.

Figure 5 shows the $\rho$-$T$ characteristics of the Ce(O,F)Sb(S,Se)$_2$ with $u = 0.5$, $x = 1.5$ single crystal. The electrical resistivity increased with decrease of temperature, implying its insulating behavior. The electrical resistivity reached to the measurement limit of resistance at 197 K. The electrical resistivity at 300 K is comparable to that of CeO$_{0.9}$F$_{0.1}$SbSe$_2$ polycrystalline samples (approximately 1 kΩcm) [22], but lower than Ce(O,F)SbS$_2$ single crystals (above 10 kΩcm) [8]. Therefore, the substitution of Se into S and/or F into O can enhance electron conductivity.

In contrast, magnetization behavior in the Ce(O,F)Sb(S,Se)$_2$ single crystals changes with nominal F-contents ($u$) and Se-contents ($x$) as shown in Figure 6. The $M$-$T$ characteristics without F-doping ($u = 0$) shows no magnetic transition, on the other hand, those of F-doped exhibit the anomaly in around 6 K. It suggests that a magnetic ordering exists at the ordering temperature ($T_m$). The appearance of this magnetic ordering is similar to that was found in the $BiS_2$-based and $SbS_2$-based compounds [6-8,23-25]. This magnetic behavior is attributed to the Ce(O,F) layers. Systematic investigation of Ce(O,F)$BiS_2$, Ce(O,F)$SbS_2$ and Ce(O,F)Sb(S,Se)$_2$ single crystals are therefore necessary to clarify the correspondence between the crystal structure and the behavior of magnetization.

Finally, PES spectra of Ce(O,F)Sb(S,Se)$_2$ with $u = 0.5$, $x = 1.5$ single crystal in energy gap region was measured for exploring the topological state. Figure 7 shows the excitation-energy dependence of the valence band PES spectra of the Ce(O,F)Sb(S,Se)$_2$ single crystal. Even at surface sensitive measurements using the low excitation energies, PES spectra do not show any signature of metallic density of states derived from the topological surface state in the energy gap region. It suggests that Ce(O,F)Sb(S,Se)$_2$ with $u = 0.5$, $x = 1.5$ single crystals do not behave as a topological insulator, against the expectation from X-ray structural analysis. Although our results of Ce(O,F)Sb(S,Se)$_2$ single crystals show no evidence of topological insulator, an increase of electrical conductivity in Ce(O,F)Sb(S,Se)$_2$ would be a potential to find a new topological insulator.

## 4. Conclusion

$Rn$(O,F)Sb(S,Se)$_2$ ($Rn$ = La, Ce) single crystals were successfully grown by using a CsCl/KCl flux. In the Ce(O,F)Sb(S,Se)$_2$ with $u = 0.5$, $x = 1.5$ single crystals, the X-ray structural analysis was carried out. The analysis revealed that Se atoms have selectively occupied to the in-plane site of the Sb-SSe layer, and then SbSe-plane was formed in Ce(O,F)Sb(S,Se)$_2$ structure. The XAFS analysis showed that the chemical state of Ce is only trivalent in Ce(O,F)Sb(S,Se)$_2$ with $u = 0.5$, $x = 1.5$ single crystals. And they did not show the topological state in energy gap region.

## Acknowledgments

The XAFS experiments were conducted at the BL5S1 of Aichi Synchrotron Radiation Center, Aichi Science & Technology Foundation, Aichi, Japan (Experimental No.201704041).

Table I. Heat treatment temperature ($T_{max}$), nominal F ($u$) and Se ($x$) composition, the analytical F and Se composition in the obtained single crystals. The analytical F and Se composition were normalized by total O+F = 1.0 and S+Se = 2.0 amounts, respectively.

| | | Nominal composition in $RnO_{1-u}F_uSbS_{2-x}Se_x$ | | Analytical composition | | | |
|---|---|---|---|---|---|---|---|
| | | | | O+F=1 | | S+Se=2 | |
| $Rn$ | $T_{max}$ (°C) | $u$ | $x$ | O | F | S | Se |
| Ce | 900 | 0 | 1.0 | 1.00 | 0 | 1.24±0.04 | 0.76±0.04 |
| | 900 | 0.5 | 1.0 | 0.7±0.1 | 0.3±0.1 | 1.16±0.06 | 0.84±0.06 |
| | 800-900 | 0 | 1.5 | --- | | | |
| | 800 | 0.5 | 1.5 | 0.8±0.1 | 0.2±0.1 | 0.93±0.01 | 1.07±0.01 |
| La | 900 | 0.5 | 1.0 | 1.00 | 0 | 1.47±0.03 | 0.53±0.03 |
| | 800-900 | 0.5 | 1.5 | --- | | | |

---:No $Rn$(O,F)Sb(S,Se)$_2$ single crystals were obtained.

Table II. Crystallographic data for the Ce(O,F)Sb(S,Se)$_2$ with $u = 0.5$, $x = 1.5$.

| | |
|---|---|
| Structural formula | Ce$_2$O$_{1.52}$F$_{0.48}$Sb$_{1.90}$S$_2$Se$_2$ |
| Formula weight | 765.85 |
| Crystal dimensions (mm) | 0.22 × 0.11 × 0.02 |
| Crystal shape | Platelet |
| Crystal system | Monoclinic |
| Space group | $P2_1/m$(No. 11) |
| $a$ (Å) | 4.121(7) |
| $b$ (Å) | 4.109(7) |
| $c$ (Å) | 13.233(15) |
| $\beta$ (°) | 97.94(7) |
| $V$ (Å$^3$) | 221.9(6) |
| $Z$ | 1 |
| $d_{calc}$ (g/cm$^3$) | 5.731 |
| Temperature (K) | 293 |
| $\lambda$ (Å) | 0.71073 (MoK$\alpha$) |
| $\mu$ (mm$^{-1}$) | 24.367 |
| Absorption correction | Empirical |
| $\theta_{max}$ (°) | 32.498 |
| Index ranges | -6<$h$<6, -5<$k$<6, -18<$l$<19 |
| Total reflections | 2235 |
| Unique reflections | 817 |
| Observed [$I \geq 2\sigma(I)$] | 374 |
| $R_{int}$ for all reflections | 0.3237 |
| No. of variables | 30 |
| $R_1/wR_2$ [$I \geq 4\sigma(I)$] | 0.1441/0.3296 |
| $R_1/wR_2$ (all data) | 0.1895/0.3699 |
| GOF on $F_o^2$ | 1.050 |
| Max./Min. residual density (e$^-$/Å$^3$) | 8.299 / -4.794 |

Table III. Atomic coordinates for the Ce(O,F)Sb(S,Se)$_2$ with $u = 0.5$, $x = 1.5$.

| Site | S.O.F | x/a | y/b | z/c |
| --- | --- | --- | --- | --- |
| Ce | 1.00 | 0.2060(5) | 1/4 | 0.40178(16) |
| Sb | 0.95(2) | 0.6160(15) | 3/4 | 0.1232(3) |
| S | 1.00 | 0.665(2) | 3/4 | 0.3068(6) |
| Se | 1.00 | 0.0851(12) | 1/4 | 0.1360(4) |
| O | 0.76(Fix) | 0.259(5) | 3/4 | 0.5000(15) |
| F | 0.24(Fix) | 0.259(5) | 3/4 | 0.5000(15) |

**Figure captions**

Figure 1. Crystal structure of $Rn$(O,F)Sb$Ch_2$.

Figure 2. Typical SEM image of Ce(O,F)Sb(S,Se)$_2$ single crystal.

Figure 3. XRD pattern of well-developed plane of Ce(O,F)Sb(S,Se)$_2$ with $u = 0.5$, $x = 1.5$ single crystal.

Figure 4. Ce L$_3$-edge, XAFS obtained at room temperature for Ce(O,F)Sb(S,Se)$_2$ with $u = 0.5$, $x = 1.5$ single crystals, Ce$_2$S$_3$, and CeO$_2$.

Figure 5. Resistivity-temperature ($\rho$–$T$) characteristics for the Ce(O,F)Sb(S,Se)$_2$ with $u = 0.5$, $x = 1.5$ single crystal.

Figure 6. Temperature ($T$) dependence of magnetization ($M$) under zero-field cooling (ZFC) and field cooling (FC) with an applied field ($H$) of 10 Oe parallel to the $c$-axis for the single crystals of Ce(O,F)Sb(S,Se)$_2$ with (a) $u = 0$, $x = 1.0$, (b) $u = 0.5$, $x = 1.0$ and (c) $u = 0.5$, $x = 1.5$.

Figure 7. Excitation photon-energy (hν) dependence of the valence band PES spectra of Ce(O,F)Sb(S,Se)$_2$ with $u = 0.5$, $x = 1.5$ single crystal.

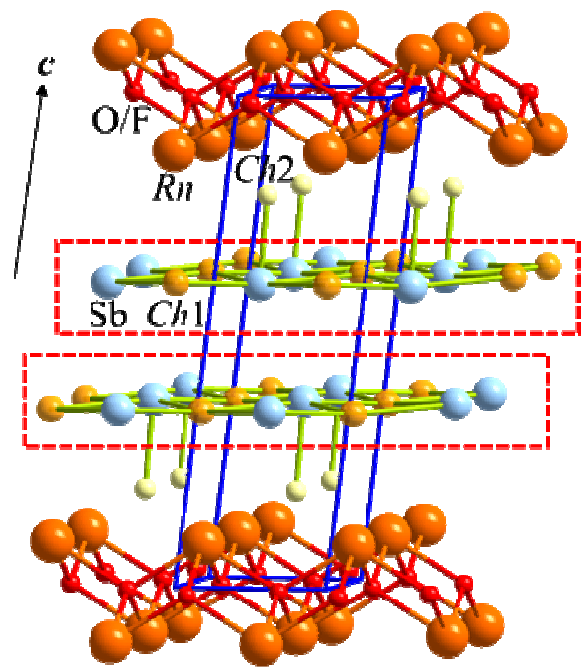

**Figure 1**

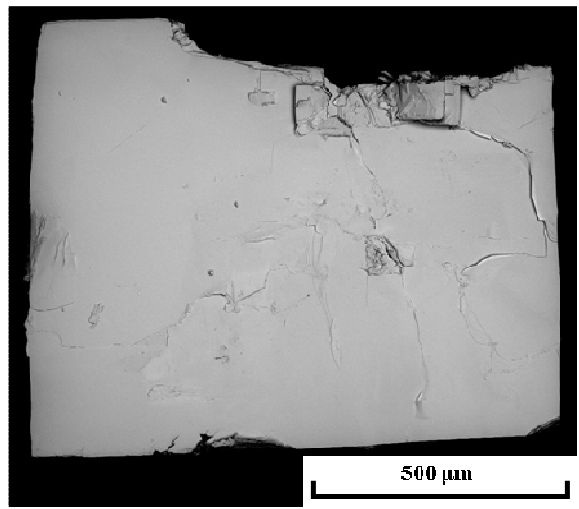

**Figure 2**

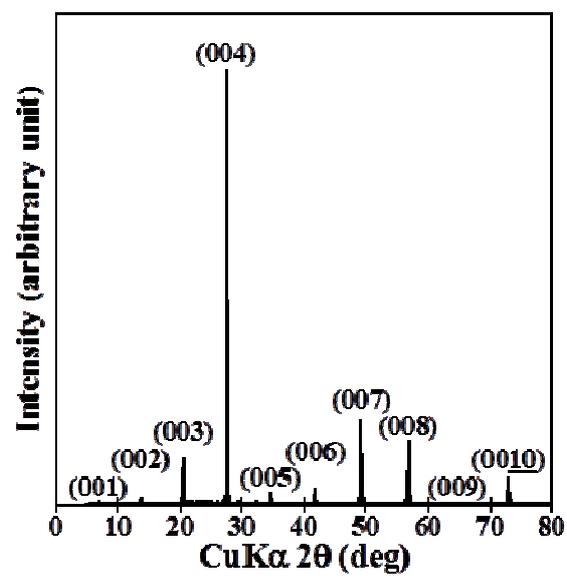

**Figure 3**

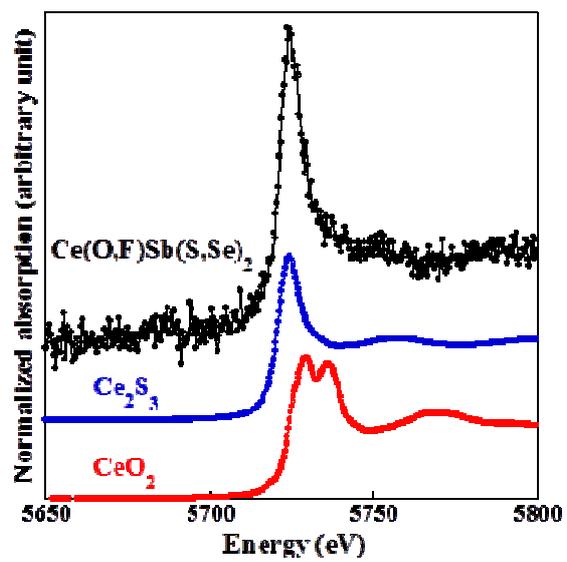

**Figure 4**

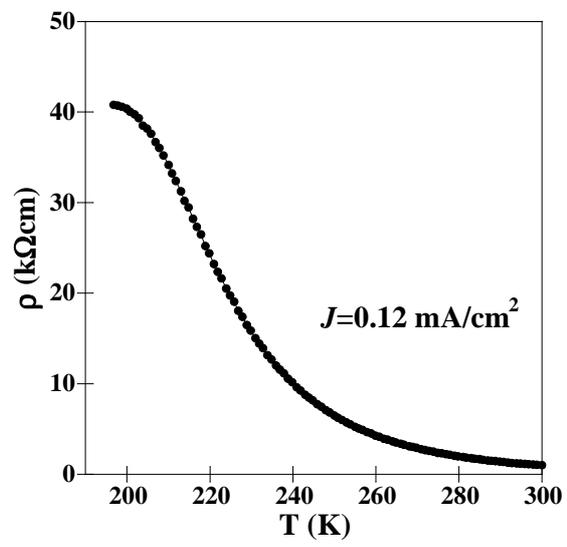

**Figure 5**

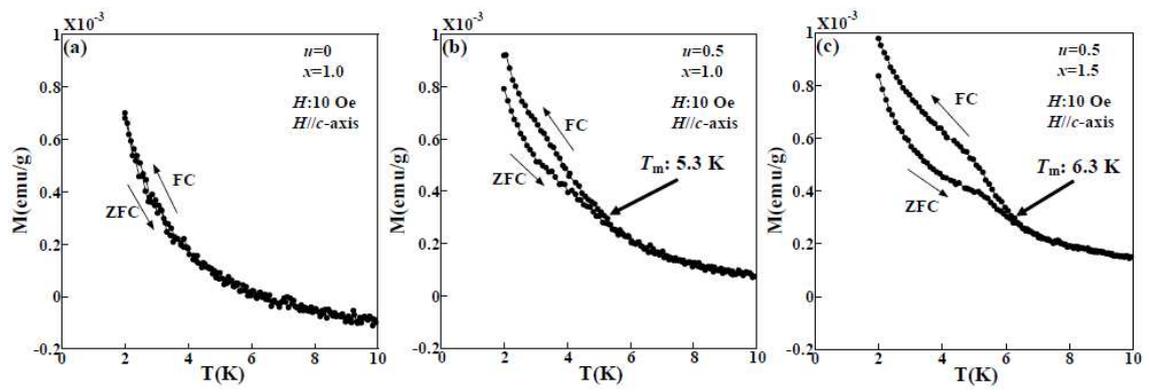

**Figure 6**

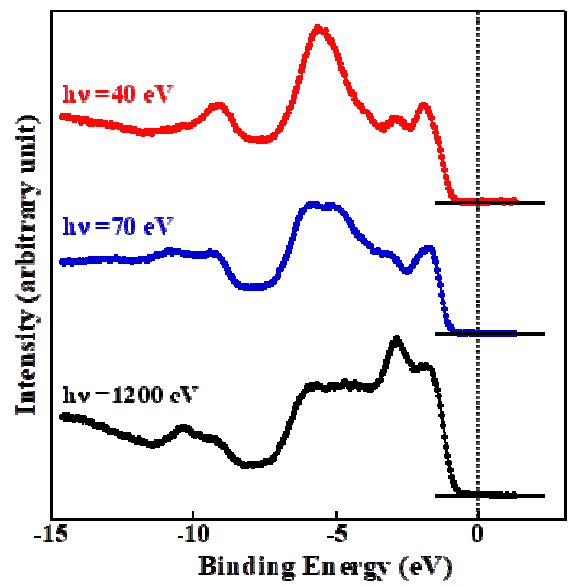

**Figure 7**